\documentclass[aps,preprint]{revtex4} \usepackage{amsmath,amssymb,amsfonts,graphicx,hyperref} 
\newcommand{\eeq}{\end{equation}} \newcommand{\bea}{\begin{eqnarray}} \newcommand{\eea}{\end{eqnarray}}

\begin{document}

\title{On Effective Spacetime Dimension in the Ho\v{r}ava-Lifshitz Gravity}

\author{G. Alencar} \email{geova@fisica.ufc.br}\address{Departamento de F\'{i}sica, Universidade Federal do Cear\'a, Caixa Postal
6030, CEP 60455-760, Fortaleza, CE, Brazil}

\author{V. B. Bezerra} \email{valdir@fisica.ufpb.br}\address{Departamento de F\'isica. Universidade Federal da Para\'iba, Caixa Postal
5008.\\CEP 58059-970. Jo\~ao Pessoa-PB, Brazil.}

\author{M. S. Cunha} \email{marcony.cunha@uece.br} \address{Grupo de F\'isica Te\'orica (GFT), Universidade Estadual do Cear\'a-UECE,
CEP 60714-903, Fortaleza-CE, Brazil}

\author{C. R. Muniz} \email{celio.muniz@uece.br}\address{Grupo de F\'isica Te\'orica (GFT). Universidade Estadual do Cear\'a,
Faculdade de Educa\c c\~ao, Ci\^encias e Letras de Iguatu. Iguatu-CE, Brazil.}

\begin{abstract} In this manuscript we explicitly compute the effective dimension of spacetime in some backgrounds of
Ho\v{r}ava-Lifshitz (H-L) gravity. For all the cases considered, the results are compatible with a dimensional reduction of the
spacetime to $d+1=2$, at high energies (ultraviolet limit), which is confirmed by other quantum gravity approaches, as well as to
$d+1=4$, at low energies (infrared limit). This is obtained by computing  the free energy of massless scalar and gauge fields. We find
that the only effect of the background is to change the proportionality constant between the internal energy and temperature.
Firstly, we consider both the non-perturbative and perturbative models involving the matter action, without gravitational sources but
with manifest time and space symmetry breaking, in order to calculate modifications in the Stephan-Boltzmann law. When gravity is
taken into account, we assume a scenario in which there is a spherical source with mass $M$ and radius $R$ in thermal equilibrium with
radiation, and consider the static and spherically symmetric solution of the H-L theory found by Kehagias-Sfetsos (K-S), in the weak
and strong field approximations. As byproducts,  for the weak field regime, we used the current uncertainty of the solar radiance
measurements to establish a constraint on the $\omega$ free parameter of the K-S solution. We also calculate the corrections, due to
gravity, to the recently predicted attractive force that black bodies exert on nearby neutral atoms and molecules. \vspace{0.75cm}
 \end{abstract}


\maketitle

\section{Introduction}

The Ho\v{r}ava-Lifshitz (H-L) gravity is an alternative proposal to quantum gravity theory, recently presented in the literature,
which is power-counting renormalizable \cite{horava1,horava2,horava3}. The cost to ensure this renormalizability is to ignore the
local Lorentz symmetry incorporated in Einstein's theory of gravitation and to consider different kinds of spatial and temporal
scaling at very short distances, {\it i.e.}, in the ultraviolet (UV) regime, with this symmetry accidentally emerging in the opposite
situation, namely, at the the infrared (IR) regime \cite{visser}. According to this novel approach, at the UV regime occurs an
anisotropic scaling, in such a way that time transforms as $t\rightarrow b^zt$ and space as $x^{i}\rightarrow bx^{i}$, where $z$ is a
dynamical critical exponent that goes to unity at large distances, which means that the validity of General Relativity is restored, at
the IR scale.

The possible values of the critical exponent $z$ establishes a connection between the H-L theory and the number of dimensions that the
Universe must possess at all scales. In fact, Ho\v{r}ava \cite{horava3} calculates the dimension seen by a diffusion process, the
spectral dimension $d_s$, as being equal to 2 at the UV scale, which is a value compatible with other approaches of quantum gravity
\cite{carlip,sehara,Celio}). At the opposite scale, which means at the IR regime, the spectral dimension is $d_s=4$ \cite{horava3}.
The spectral dimension as a continuous function of time which links these two amounts, in the context of the H-L theory, was obtained
in \cite{brito}. A general discussion about the dependence of the dimensionality with the scale (``evolving" or ``vanishing"
dimensions), in context of several theories, can be found in \cite{Dejan}. Ho\v{r}ava still presents \cite{horava3} a heuristic
scaling arguments based on the free energy of a system of free massless fields to show that the effective number of topological
dimensions is again equal to 2, since that the free energy found is proportional to the square of temperature.

Studies concerning H-L gravity were made in several contexts, as in the analysis of the Casimir effect \cite{petrov, muniz} and in the
study of spacetime stability \cite{huang}, as well as in some astrophysical and cosmological models \cite{Alexandre,son,fonseca}, including those which predict a
multiverse \cite{faizal}. It was shown that H-L gravity is weaker than General Relativity even at high energies, generating a bouncing
cosmology and an accelerating universe, naturally incorporating dark energy \cite{mukohyama,corrado}. From the point of view of
Astrophysics, some static and spherically symmetric black hole-type solutions of H-L gravity were found, and the simplest one is that
proposed by Kehagias and Sfetsos (K-S) \cite{kehagias}, whose IR limit gives us the Schwarzschild solution. Other exact spherically
symmetric solutions for the H-L gravity can be found in \cite{pope,park,kiritsis}.

The connection between the phenomenology involving gravity and thermodynamics was studied in the early seventies of past century
\cite{Hawking1,Hawking2}, and reexamined more recently \cite{Jacobson,Verlinde}. In this context, Black Holes seem to be the objects
of the universe in which these connections become stronger, and any modifications in the laws derived from thermodynamics due to
gravitational effects must consider these structures, necessarily. In this scenario, black body radiation is emblematic, in view of
the possible role played by this phenomenon in the construction of a quantum theory of gravity.

In this paper, we calculate explicitly the number of effective topological dimensions of the Universe by analyzing modifications in
the laws of the black body radiation due to the H-L theory, at both UV and IR scales with and without gravitation sources. It is
supposed that the thermal radiation comes from the dynamics of a free massless scalar field.  We then calculate the Helmholtz
free-energy associated with the field for the cases under consideration. Firstly, the non-perturbative model, in which, due to the
spacetime anisotropy required by H-L theory, the spatial derivatives of the scalar field enter in the action with higher orders than
the temporal one. In what follows, we will take into account the perturbative model, where, besides these derivatives, one adds more a
term which depends on the first order spatial derivative. We use the models presented in \cite{petrov} to obtain the dependence of the
Helmholtz free-energy with the temperature and to  determine the effective number of spacetime dimensions. Next we consider a
spherical gravitational source, by considering the weak and strong field approximations of the K-S solution, in both IR and UV
regimes. Finally, for the weak field regime, we will obtain constraints on the K-S $\omega$ parameter from the uncertainty in the
measurements of the solar radiance. We will also calculate corrections, due to both General Relativity and H-L theories, to the
thermal shift in the energy of the fundamental state of the hydrogen atom, known as dynamic Stark effect, obtaining the anomalous
force of attraction upon it due to the black body radiation.

The paper is organized as follows. In section 2, we calculate the modifications in the laws of the black body radiation in the context
of the H-L theory without considering the presence of a gravitational field. In section 3, we do similar calculations to the previous
section, but now taking into account the presence of a gravitational field. In section 4 we obtain the force of attraction corrected
by effects due to classical and H-L gravity. Finally, in section 5 we present the concluding remarks.

\section{Black body Radiation Without Gravitational Sources}

\subsection{Non-perturbative Model} In the H-L theory, the violation of the Lorentz symmetry at high energies comes from the different
forms of scaling time and space. This feature generates changes in the action of a free massless scalar field with respect to that one
described in Minkowski spacetime. In this sense, the more general action for such a field can be written as \begin{equation} S=\int
dtd^dx\frac{1}{2}\left[(\partial_t\phi)^2-\sum_{s=1}^zg_s\left(\partial^{i_1}\partial^{i_2}...\partial^{i_s}\right)\phi\left(\partial_{i_1}\partial_{i_2}...\partial_{i_s}\right)\phi\right].
\end{equation}

In the non-perturbative model is that in which the Eq. (1) only contains the terms where $s=z$ and one constant $g_z$, which we will
write as $g_z=\ell^{2(z-1)}$, where $\ell$ is a characteristic length. Thus the field equation, according to this model, is given by
\begin{equation}\label{1} \partial_t^2\phi-\ell^{2(z-1)}\nabla^{2z}\phi=0 \end{equation} Considering an oscillatory field of the form
$\phi(\vec{x},t)=C \exp{(i\vec{k}.\vec{x}-i\varpi t)}$, we get $\varpi=\ell^{(z-1)}k^{z}$, where $k=\sqrt{k_x^2+k_y^2+k_z^2}$.

The black body radiation is described by the Stephan-Boltzman law, which can be found from the Helmholtz free-energy density,
expressed by \cite{Feynman} \begin{equation}\label{3}
f_{bb}(T)=2k_BT\int\frac{d^3\textbf{k}}{(2\pi)^3}\log\{{(1-e^{-\widetilde{\beta}k^z})}\} , \end{equation} where
$\widetilde{\beta}=\ell^{z-1}/k_BT$. The factor 2 was introduced to account for the two modes of polarization of the electromagnetic
waves. When $z=1$, namely, in the low-energies domain, this free-energy density is \begin{equation}\label{4}
f_{bb}(T)=-\frac{\pi^2(k_BT)^4}{45}. \end{equation} which leads to the law of Stephan-Boltzman, from which we obtain the energy
density of the black body, given by \begin{equation} u_{bb}(T)=-T^2\frac{\partial (f_{bb}/T)}{\partial T}=\frac{\pi^2(k_B T)^4}{15},
\end{equation} as it should be in the IR limit and in absence of a gravitational field.

On the other hand, if we put $z=3$ in Eq. (\ref{3}), which corresponds to high-energies scale, the free-energy density will be now
given by \begin{equation}\label{5} f_{bb}(T)=-u_{bb}(T)=-\frac{(k_BT)^2}{18\ell^2}. \end{equation}

It is known that the of Stephan-Boltzman law in a $d$-dimensional flat space yields a black body energy density proportional to
$T^{d+1}$ \cite{grassi,alnes}. Thus, by the above expression, we can see that the effective dimension of the spacetime at
high-energies scales is $d+1 = 2$, according to the Horava-Lifshitz theory. This fact will not change when we take into account the
gravitational field. Such specific dimensional reduction at UV scale is  compatible with other approaches of quantum gravity
\cite{carlip,sehara}. In fact, this dependence with the square of temperature  works for any d-dimensional flat space, provided that
$z=d$ in order to have the renormalizability of the theory. The measure of the integral (\ref{3}) is
$[\pi^{-d/2}2^{-d}d/\Gamma(d/2+1)]k^{d-1}$, where $\Gamma(x)$ is the Gamma function and the black body energy density is exactly given
by \begin{equation}\label{8} u_{bb}(T)=\frac{2^{-d}\pi^{(4-d)/2}\ell^{1-d}}{3\Gamma(d/2+1)}(k_BT)^2. \end{equation} For an arbitrary
temperature, a numerical analysis of the above equation reveals that the energy density has a minimum for $d=3$ when
$\ell\approx0.2\ell_P$, with $\ell_P$ being the Planck length, and thus extending the UV domain of the H-L theory to trans-Planckian
scales \cite{koh}.

\subsection{Perturbative Model}

Regarding the perturbative model, which consists of adding to the action of the non-perturbative model a term proportional to the
first order spatial derivatives, in such a way that at IR scales the higher order derivatives are neglected, the dispersion relation
is given by $\varpi=k\sqrt{1+(\ell k)^{2(z-1)}}$\cite{petrov}. In this case, the Helmholtz free-energy of the black body radiation
will be, therefore \begin{equation}\label{7} f_{bb}(T)=2k_BT\int \frac{d^3\textbf{k}}{(2\pi)^3}\log \bigg\{1-\exp\left[{-\beta
k\sqrt{1+(\ell k)^{2(z-1)}}}\right]\bigg\}, \end{equation} where $\beta=(k_BT)^{-1}$. It becomes evident that, in the IR regime,
namely when $z=1$, the free-energy has the same dependence with the temperature as the one obtained in 4-dimensional Minkowski
spacetime.

At UV scale, specifically for $z=3$, the integration of (\ref{7}) is not analytically soluble, and thus we have depicted, in Fig. 1,
some values of $f_{bb}$ as a function of $t=k_B T$, for $\ell=1$, obtained numerically. In the same graph, we fit the curve that best
represents this function, which was found to be $f_{bb}\simeq -0.049117 t^2$. \begin{figure}[!ht] \centering
\includegraphics[scale=0.9]{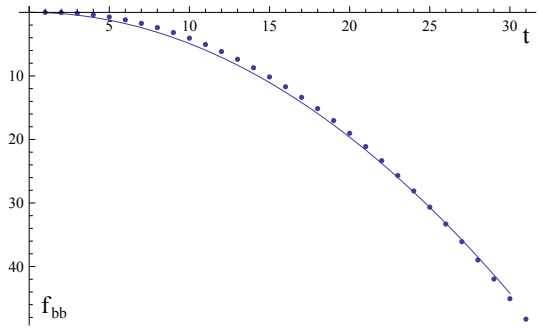} \caption{Helmholtz free-energy density of the black body radiation as a
function of $t=k_B T$, in the perturbative model of the H-L theory.} \end{figure}

Then, due to the quadratic dependence of the black body density energy with the temperature, the same conclusions valid for the
non-perturbative model are applied here, by considering $z=3$.

\section{Black body radiation due to a spherical gravitational source}

The K-S black hole solution is given by the metric \cite{kehagias} \begin{equation}\label{8}
ds^2=f_{KS}(\rho)dt^2-\frac{d\rho^2}{f_{KS}(\rho)}-\rho^2d\Omega^2, \end{equation} where \begin{equation}\label{9}
f_{KS}(\rho)=1+\omega \rho^2\left(1-\sqrt{1+\frac{4M}{\omega \rho^3}}\right), \end{equation} with $\omega$ being a free parameter of
the H-L theory and $\rho$ a radial coordinate. It is easy to note from (\ref{9}) that, in the coordinates system used, the K-S
solution has two event horizons, one at $\rho_{out}=M(1+\sqrt{1-1/2\omega M^2})$ and another at $\rho_{int}=M(1-\sqrt{1-1/2\omega
M^2})$.

The K-S solution represents a black hole if $\omega M^2\geqslant 1/2$, while for $0 < \omega M^2 < 1/2 $ it is a singularity without
event horizons (naked singularity). In the limit $\omega \rightarrow \infty$, where $\omega$ is the H-L parameter, it reduces to the
Schwarzschild solution.

\subsection{Weak-field approximations}

When $\omega\rightarrow\infty$ (IR limit), the metric coefficients of Eq. (\ref{8}) tend to those of Schwarzschild solution, i.e.,
$f_{KS}(\rho)\rightarrow f_{Sch}(\rho)=1-2M/\rho$, which corresponds to the vacuum solution generated by a static spherical source.
Thus, this result is consistent with the validity of the General Relativity at IR scale.

We will take into account a static background gravitational field described by the K-S solution, whose metric contains an irremovable
singularity at $\rho=0$. We suppose that the radius $R$ of the spherical body is much larger than the radii of the event horizons. In
other words, we will work in a weak field approximation.

We already have emphasized that our approach will occur close to the IR limit ($z\approx1$, $\omega\gg1$). Thus, we will make an
expansion in the metric coefficient given by Eq. (\ref{9}) so that \begin{equation}\label{10}
f_{KS}(\rho)\approx1-\frac{2M}{\rho}+\frac{2M^2}{\omega\rho^4}. \end{equation} The last term in (\ref{10}) is what effectively
incorporates the properties of Ho\v{r}ava-Lifshitz gravity. We can note at a glance that the attractive post-Newtonian potential is
corrected by the repulsive term $M^2/\omega\rho^4$.

For our proposes, it is convenient the use of isotropic coordinates. Thus, it is necessary to find a transformation in the radial
coordinate, $\rho=\rho(r)$, so that the K-S solution has the form $ds^2=g(r)dt^2-h(r)(dr^2+r^2d\Omega^2)$. Following \cite{iorio1}, a
suitable form of the metric is given by \begin{eqnarray}\label{12} ds^2=\left(1-\frac{2M}{r}+\frac{2M^2}{\omega
r^4}\right)dt^2-\left(1-\frac{2M}{r}+\frac{M^2}{2\omega r^4}\right)^{-1}(dr^2+r^2d\Omega^2). \end{eqnarray} where we have introduced
the isotropic coordinates in order to use the rectangular coordinates\cite{muniz}.

Now, let us calculate the eigenfrequencies of the massless scalar field at the surface $r=R$ of the spherical gravitational source, in
both the weak field approximation and IR regime of the K-S solution. To do this, we will rewrite the metric (\ref{12}) in the
following form \begin{equation}\label{13} ds^2=(1+2b\Phi_0')dt^2-(1-2\Phi_0')(dx^2+dy^2+dz^2), \end{equation} where
$b=1-\frac{3M}{4\omega R^3}$ and $\Phi_0'=-\frac{M}{R}+\frac{M^2}{4\omega R^4}$.

The general covariant Klein-Gordon equation for the massless scalar field in the spacetime given by equation (\ref{13}) is written as
\begin{equation}\label{15}
\frac{1}{\sqrt{-g}}\partial_{\mu}(\sqrt{-g}g^{\mu\nu}\partial_{\nu})\Psi=[1-2(b+1)\Phi_0']\partial_t^2\Psi-\ell^{2(z-1)}\nabla^{2z}\Psi=0.
\end{equation}

Considering plane wave solutions we can find that the eigenfrequencies $\varpi$ are given by the expression
$\varpi=\ell^{z-1}\left(1-\frac{2M}{R}+\frac{5M^2}{4\omega R^4}\right)k^{z}$. The Helmholtz free-energy density of the black body
radiation is \begin{equation}\label{18} f_{bb}(T)=2k_BT\int
\frac{d^3\textbf{k}}{(2\pi)^3}\log{\left[1-e^{-\widetilde{\beta}k^{z}}\right]}, \end{equation} where the factor 2 once more was
introduced to account for the two polarization modes of the electromagnetic field and $\tilde{\beta}=\ell^{z-1}
\frac{[1+\Phi_0'(b+1)]}{k_BT}$. Comparing Eq. (\ref{18}) with our previous result, obtained in Eq. (\ref{3}), we get \cite{muniz} for
$z=1$ \begin{equation}\label{22} u_{bb}(T)=\frac{\left(1+\frac{6M}{R}-\frac{15M^2}{4\omega R^4}\right)\pi^2(k_BT)^4}{15}.
\end{equation} On the other hand, if we put $z=3$ in Eq. (\ref{18}), which corresponds to high-energies scale, we get \begin{equation}
\label{20} u_{bb}(T)=\left(1+\frac{6M}{R}-\frac{15M^2}{4\omega R^4}\right)\frac{(k_BT)^2}{18\ell^2}. \end{equation} Notice that,
turning off the gravity, equations (\ref{22}) and (\ref{20}) recover our previous results obtained in section II. Therefore for the
weak field approximation we again obtain that, for $z=1$, the spacetime effective dimension is $4$. For $z=3$ we get the expected
dimensional reduction to $d=1$.

It is convenient to define an effective temperature given by \begin{equation}\label{23}
T_{eff}=\left(1+\frac{3M}{2R}-\frac{15M^2}{16\omega R^4}\right)T. \end{equation} We note that, if one considers just up to $M/R$
corrections ({\it i.e.}, due to Einstein's gravity in its post-Newtonian limit), the intensity of the black body radiation is slightly
augmented. With respect to the Sun, for example, this represents an increment of 0.0012 per cent comparing to a black body one at same
temperature in Minkowski space. On the other hand, the correction term associated to the H-L gravity represents only a very low
attenuation of that radiation. Based on the current uncertainties in measurements of the solar radiance ($\approx 0.02$
\cite{Thuillier}), we can estimate a lower limit for the K-S parameter as being $\omega\geq3\times10^{-31}cm^{-2}$.

\subsection{Strong-field approximations}

The K-S metric at the high-energies scale, in which $\omega\ll1$, is given by \cite{dwornik} \begin{equation}\label{26}
ds^2=\left[1-2\left(\omega M\rho\right)^{1/2}\right]dt^2-\left[1-2\left(\omega M\rho\right)^{1/2}\right]^{-1}d\rho^2-\rho^2d\Omega^2.
\end{equation} We have to find an isotropic reference frame such that $r=r(\rho)$ in order to getting \begin{equation}\label{27}
ds^2=g(r)dt^2-f(r)^2(dr^2+r^2d\Omega^2), \end{equation} where $f(r)=\rho/r$ and \begin{equation}\label{28}
f(r)dr=\frac{d\rho}{\sqrt{1-2(\omega M\rho)^{1/2}}}. \end{equation} Integrating this and choosing an appropriate constant of
integration, we have \begin{equation}\label{29} r(\rho)=M\left(\frac{\sqrt{1-2(\omega M\rho)^{1/2}}-1}{\sqrt{1-2(\omega
M\rho)^{1/2}}+1}\right)^2, \end{equation} and \begin{equation}\label{30} \rho(r)=\frac{r^2}{\omega M(M-r)^2}, \end{equation} such that
$g(r)=1-2r/(M-r)$ and $f(r)=r/\omega M(M-r)^2$.

The metric (\ref{27}) can be written, up to $\mathcal{O}(r/M)$, as \begin{equation}\label{31}
ds^2\thickapprox\left(1-\frac{2r}{M}\right)dt^2-\frac{r^2}{\omega^2 M^6}\left(1+\frac{4r}{M}\right)(dr^2+r^2d\Omega^2). \end{equation}
This metric must be equivalent to that one of Eq. (\ref{26}) in the limit of $\omega\rightarrow 0$, provided that the limits
$r\rightarrow 0$ and $(r/\omega M^3)\rightarrow 1$ are satisfied.

Let us now write the massless scalar covariant Klein-Gordon equation in the spacetime given by Eq.(\ref{31}), taking $r=R$. Thus, we
get \begin{equation} \left(1+\frac{6R}{M}\right)\frac{R^2}{\omega^2M^6}\partial_t^2\Psi-\ell^{2(z-1)}\nabla^{2z}\Psi=0, \end{equation}
where $\ell$ is the characteristic length \cite{petrov}.

Writing the dispersion relation for the massless scalar field using the ansatz $\Psi=C\exp(i{\bf k}\cdot{\bf x}-i\varpi t)$, we obtain
$\varpi=\frac{\ell^{z-1}\omega M^3}{R} \left(1+\frac{6R}{M}\right)^{-\frac{1}{2}}k^z$. In this case, the Helmholtz free-energy density
is given by \begin{equation}\label{33} f_{bb}(T)=2k_BT\int
\frac{d^3\textbf{k}}{(2\pi)^3}\log{\left[1-e^{-\widetilde{\beta}k^z}\right]}, \end{equation} where $\tilde{\beta}=\ell^{z-1}\omega
M^3(1+6R/M)^{-\frac{1}{2}}/(k_BTR)$. Following the same steps as in the previous sections we get for $z=3$ \begin{equation}\label{34}
u_{bb}(T)=\frac{R}{18\ell^{2}\omega M^3}\left(1+\frac{6R}{M}\right)^{\frac{1}{2}}(k_BT)^2. \end{equation} and for $z=1$
\begin{equation}\label{ff} u_{bb}(T)=\frac{R}{\omega M^3}\left(1+\frac{6R}{M}\right)^{\frac{1}{2}}\frac{\pi^2(k_BT)^4}{15},
\end{equation} which are the expected results for the UV and IR regimes.

\section{The ``black body force''}

The universal properties of the black body radiation depend on its temperature. Such radiation has always been thought to have a
repulsive effect on other bodies situated near it, via momentum transference of the emitted photons. Studying the shift in the
spectral lines of the hydrogen atom placed in a thermal bath, it was shown that the black body radiation exerts another quite novel
and counter-intuitive attractive force on nearby neutral atoms and molecules, which is even more intense than the repulsive radiation
pressure or the gravity generated by the black body. This interaction was described by the first time in \cite{Marte} and according to
their authors, the attractive potential associated to the force exerted on a hydrogen atom in its fundamental state, placed at a
distance $r$ from the center of a spherical black body of radius $R$ that emits radiation at a temperature $T$ is, in natural units,
\begin{equation}\label{24} V_{bb}(r)=-\frac{3\pi^2 (k_BT)^4}{10 (e m_e)^3}\left(1-\frac{\sqrt{r^2-R^2}}{r}\right), \end{equation}
where $e$ and $m_e$ are the electric charge and the mass of the electron, respectively. The last factor in the parenthesis is the
solid angle through which the atom sees the source.

It is worth point out that a modification in the solid angle due to the spacetime curvature takes place here. The solid angle that
appears in the equation above depends on the radial distances, which are defined now according to the $g_{rr}$ coefficients that
appear in both interior and exterior solutions, with the latter given by Eq. (\ref{12}). In the case of the General Relativity, the
simplest spherically symmetric interior solution is given by the Oppenheimer-Snyder metric \cite{Snyder}, in which
$g_{rr}=\left(1-\frac{r^2}{R_0^2}\right)^{-1}$, where $R_0=3c^2/8\pi G\rho_0$, with $\rho_0$ being the density of the ideal fluid that
constitutes the body. For the Sun, whose average density is $\rho_0\sim$ $10^{3}$kg$/$m$^3$, $R_0\sim 10^{26}$m, thus we have that
$r\leq R$ implies $r\ll R_0$, and therefore $g^{in}_{rr}\simeq 1$. We will assume the same in the H-L theory since that in this
approach the gravitational interaction is weaker than those one of Einstein's theory.

In the weak-field approximation that we are using here, the solid angle seen by the atom will be modified through the radial distances
in terms of which it is expressed as \begin{equation}\label{24.1} r\rightarrow
\overline{r}=\int_{0}^{R}\sqrt{g^{in}_{rr}}dr+\int_{R}^{r}\sqrt{g^{ex}_{rr}}dr=
r+M\ln\left(\frac{r}{R}\right)-\frac{M^2}{12\omega}\left(\frac{1}{R^3}-\frac{1}{r^3}\right). \end{equation}

The source radius $R$ remains unaltered since that $g_{rr}^{in}\simeq1$, as we have already seen. The temperature that we will
consider here is the effective one given by Eq. (\ref{23}). This will allow us to include corrections in the effect due to both
classical and quantum gravity theories, which are under consideration in this paper. We have therefore \begin{equation}\label{25}
V_{bb}(r)=-\frac{3\pi^2\left(1+\frac{6M}{R}-\frac{15M^2}{4\omega R^4}\right) (k_BT)^4}{10 (e
m_e)^3}\left[1-\frac{\sqrt{\overline{r}^2-R^2}}{\overline{r}}\right], \end{equation} where $\overline{r}$ is given by Eq.
(\ref{24.1}). Thus, the attractive force is intensified by the Einstein's gravity ($\approx 0.0012$ per cent) and attenuated by the
Ho\v{r}ava-Lifshitz gravity, independently of the atom's distance from the black body center.

 \section{Concluding remarks}

In this paper we have analyzed implications on the black body radiation which is derived from the dynamics of massless scalar and
gauge fields in two different situations, with and without a gravitational source in the framework of the Ho\v{r}ava-Lifshitz theory.
In the gravity free case, the study was based on both the non-perturbative and perturbative models for the field action, following a
recent published result \cite{petrov}. For the case with source we have supposed that the black body has mass $M$ and radius $R$
generating a static, spherically symmetric and asymptotically flat background gravity described by the Kehagias-Sfetsos solution. For
the latter analysis, it was considered  the weak and strong field approximations.

We found the Helmholtz free-energy of the black body in order to determine its gravitationally corrected energy density.  In all cases
considered we showed that, at IR scale, when the space and time symmetry breaking parameter is $z=1$, the black body energy density is
proportional to $T^4$, which corresponds to an Universe with spacetime dimensionality equal to 4. In UV regimen, when $z=3$, we found
that the black body energy density is proportional to $T^2$, compatible with an effective number of dimensions equal to $2$, which is
obtained through other quantum gravity theories. It is worth notice that, specifically for the sourceless case, we showed that the
quadratic dependence of the free-energy density with the temperature is independent of the topological dimension $d$, when we impose
$z=d$ to warrant the power-counting renormalizability of the theory at UV scale. With this analysis, we made more rigorous the
argument based on anisotropic scaling of the free energy used by Ho\v{r}ava in \cite{horava3}.

Applying the above results to the Sun, we verified that the Einstein's gravity intensifies the solar radiance of about 0.0012 per cent
as compared with the value obtained in flat spacetime. The term associated to the Ho\v{r}ava-Lifshitz gravity provides a lower bound
for the free parameter $\omega\geq 10^{-31}cm^{-2}$, based on the current uncertainties in the measurements of the solar radiance
($\approx 0.02$ \cite{Thuillier}). If we consider that the Sun is not a perfect black body, then a substantial part of this
measurement error comes from that assumption, what can contribute to improve the constraint upon $\omega$, approximating it of that
obtained via classical tests of General Relativity \cite{harko}, in which $\omega\geq10^{-27}cm^{-2}$.

Finally, regarding to the weak-field and IR regime, we found the corrections due to gravity to the attractive force on neutral atoms
that are under the action of the black body radiation, according to a recently published work \cite{Marte}. When the atom is very
close to the black body spherical surface, Eq. (29) shows that the force on the particle tends to infinity, as in the flat spacetime
case. In the general situation, our correction indicates that this force will vary with the source radius of form considerably
different from that case. And if we consider the above mentioned lower bound $\omega\geq10^{-27}cm^{-2}$, the attenuation of the black
body force caused by the Ho\v{r}ava-Lifshitz gravity cannot be greater than 0.00007 per cent.

\section*{Acknowledgements}

V.B.Bezerra and G. Alencar would like to thank CNPq for partial financial support.

\end{document}